\documentclass[journal,draftclsnofoot, onecolumn]{IEEEtran}
\usepackage{graphicx}
\usepackage{dcolumn}
\usepackage{amsmath,amsfonts}
\usepackage{amssymb,amsthm}
\usepackage{bm}

\begin{document}

\title{A fast marching method based back projection algorithm for photoacoustic
tomography in heterogeneous media}

\author{Tianren Wang and Yun~Jing %
\thanks{T. Wang and Y. Jing are with the Department of Mechanical
and Aerospace Engineering, North Carolina State University, Raleigh, NC 27695 (e-mail: yjing2@ncsu.edu)}
\thanks{Manuscript received Dec. 28, 2011.}}

\date{\today}

\maketitle

\begin{abstract}
This paper presents a numerical study on a fast marching method
based back projection reconstruction algorithm for photoacoustic
tomography in heterogeneous media. Transcranial imaging is used here
as a case study. To correct for the phase aberration from the
heterogeneity (i.e., skull), the fast marching method is adopted to
compute the phase delay based on the known speed of sound
distribution, and the phase delay is taken into account by the back
projection algorithm for more accurate reconstructions. It is shown
that the proposed algorithm is more accurate than the conventional
back projection algorithm, but slightly less accurate than the time
reversal algorithm particularly in the area close to the skull.
However, the image reconstruction time for the proposed algorithm
can be as little as $124$ ms when implemented by a GPU ($512$
sensors, $21323$ pixels reconstructed), which is two orders of
magnitude faster than the time reversal reconstruction. The proposed
algorithm, therefore, not only corrects for the phase aberration,
but can be also potentially implemented in a real-time manner.

\end{abstract}


\section{Introduction}

Various algorithms with different experimental configurations have
been
proposed~\cite{M_XU_2003}\cite{treeby2010}\cite{zhangc_2010}\cite{Niederhauser_2005}
to reconstruct the photo-acoustic tomography (PAT) image. Back
projection (BP) is the most commonly adopted one~\cite{HXU2007} due
to its simplification and universality to a wide range of sensor
configurations. The BP is based on the assumption that the speed of
sound distribution is relatively constant. This is typically
considered valid for soft tissue. For heterogeneous media, such as
for brain cerebral vascular imaging, however, traditional BP
algorithm results in poor reconstruction images. This is mainly
because of the presence of the skull, of which the acoustic
properties differ dramatically from the surrounding medium. For
example, the speed of sound in the skull is $2200-3000$ m/sec while
the speed of sound in the brain tissue is around $1540$ m/sec
\cite{JinX2008}. Such a strong heterogeneity violates the assumption
of the back projection algorithm. On the other hand, even for weakly
heterogeneous media, BP algorithm is shown to be suboptimal
\cite{JoseJ2012}.  Some modified BP methods have been proposed to
mitigate the effect of phase aberration, such as using a half-time
algorithm \cite{Anaostasio2005}. A ray-tracing method can be
incorporated to correct for the phase distortion \cite{JinX2008}.
Nevertheless, it has been so far limited to simple, three-layered
geometries.

Time reversal reconstruction has become progressively popular in
recent years with the development of high-performance computers. In
time reversal reconstruction, all the acoustic transducers are
considered as sources and the received acoustic signals are
propagated backwards in a time-reversed manner \cite{LiuZ2013}.
Time-reversal reconstruction arrives at better results as it takes
full advantages of the wave equation. However, the time reversal
reconstruction imaging method is very time consuming because the
full wave equation is solved in the time-domain by either the
finite-difference time-domain (FDTD) method or the k-space method,
therefore is less applicable for real-time imaging. Iterative image
reconstruction algorithms suffer from the similar drawback in terms
of image reconstruction speed \cite{JoseJ2012}\cite{HuangC2013}.
Although PAT images can be processed off-line, real-time PAT
provides unique opportunities, such as detection and interpretation
of brain activation patterns for online decision making in
neuroscience and clinical applications (e.g., lie detection and
pre-surgical mapping~\cite{Lachaux2007}). It also allows for
real-time monitoring of interventional procedures and drug delivery.

The goal of this study is to investigate a fast marching method
(FMM) based, non-iterative BP algorithm, which can be potentially
realized in real-time.  The FMM calculates the arrival time from a
pixel (to be imaged) to all sensors in order to account for the
phase delay due to the heterogeneity, provided that the speed of
sound distribution is known. The phase delay can be simply included
in the BP algorithm. Numerical simulations will be presented for PAT
brain cerebral vascular imaging to demonstrate the effectiveness of
the proposed algorithm. The paper is structured as follows: In Sec.
II the wave propagation model and FMM are briefly revisited; the FMM
based BP algorithm for PAT in heterogeneous media is then
introduced. Section III discusses simulation results of
photoacoustic transcranial imaging using the proposed and other
existing methods. Section IV discusses and concludes the paper.

\section{Algorithm}

\subsection{Wave propagation}

\subsubsection{Photo-acoustic wave equation}

Because the PAT initial pressure is typically on the order of $10$
kPa \cite{treeby2010}, the governing equation comprises the
linearized equations of mass, continuity, and pressure:

\begin{align}\label{Westervelt_w}
\frac{\partial}{\partial t} u(x,t) = - \frac{1}{\rho_0 (x)} \nabla p(x,t) \\
\frac{\partial}{\partial t} \rho(x,t) = - \rho_0(x)\nabla u(x,t) \\
p(x,t) = c (x)^2 \rho(x,t)
\end{align}

with the initial conditions being:

\begin{equation}\label{ini_condition_pat}
p(x,t)|_{t=0}=p_0(x), u(x,t)|_{t_0}=0,
\end{equation}

where $p$ is the acoustic pressure at time $t$ with location $x$,
$c$ is the sound speed, $u$ is the acoustic particle velocity,
$\rho$ is the acoustic density and $\rho_0$ is the ambient density.
Even though shear waves could be generated with the signal
propagates from brain tissue to the skull, it was not considered in
this study because 1: shear waves are strongly attenuated when
propagating in the skull \cite{PJWHITE2006}; 2: the sound speed of
shear waves in the skull is around $1500$ m/sec which is much slower
than the sound speed of the longitudinal wave \cite{PJWHITE2006} ,
so it will take longer time for the shear wave to arrive at the
sensor. The center frequencies are also distinct from shear and
longitudinal waves \cite{JinX2008}. The time difference and
frequency difference make it easy to discriminate the shear wave
from the longitudinal wave.

\subsubsection{K-space method}
Previous studies \cite{treeby2010}\cite{JingY2012b} have shown that the k-space
algorithm is accurate and efficient for solving the wave equation. In this
study, a k-space method based open source software K-WAVE \cite{treeby2010}  was
adopted to simulate the PAT ultrasound signal. Details of implementation of the
k-space algorithm can be found at \cite{treeby2010}.

\subsection{Fast marching method}
It has been shown \cite{twang2013}\cite{sethian2000} that FMM is an
effective method to compute the evolution of wave fronts governed by
the Eikonal equation, which is a high frequency approximation of the
wave equation (i.e., the scale of variation of the speed of sound is
assumed to be much greater than the acoustic
wavelength~\cite{modgil2009}) and reads

\begin{equation}
|\nabla T|^2 = 1/c^2,
\end{equation}

where $T$ is the wave front arrival time and $c$ is the sound speed
at a given location. For a 2D situation, starting with an initial
condition, the neighboring point with the shortest arrival time is
calculated. This repeats recursively until all the points in the
region are visited. A second-order approximation was used to
calculate the arrival time \cite{sethian2000}.

\begin{equation}
\frac{\partial T}{\partial x} =
\frac{3T(x,y)-4T(x-1,y)+T(x-2,y)}{2}.
\end{equation}

The proposed BP method makes use of the arrival times/time-of-flight
(TOF) from all pixels to each of the sensor point.  This can be
achieved by assigning each pixel point as a point source, and
calculates the corresponding TOF map. However, as can be seen in the
numerical simulation section, there are typically a lot more pixel
points than sensor points.  We therefore assigned each sensor point
as a point source, and calculated the TOF at all pixel points of
interest, since the ray path is reversible. The TOF can be
integrated into the BP method for reconstruction as shown below.

\subsection{FMM based back projection}
BP, which is also known as the Kirchhoff migration, has been widely
used in seismology imaging \cite{Schneider1978}. In the past decades, back
projection and its various modified version have been applied to biomedical
imaging applications \cite{HXU2007} . A recent study on breast ultrasound
imaging shows that the Kirchhoff migration can be improved by taking the arrival
times in a heterogeneous medium into account \cite{schmidt2011}.

In that paper, the TOF has to be calculated for a ray path from the source to
the pixel, and then to the sensor. In PAT, only one-way propagations are
present. The modified BP in combination with the FMM method is
shown below as:

\begin{equation}\label{BP}
P(\overrightarrow{r_j}) = \sum_{i=\text{receiver index}}
[p_i(t_{i,j}) - t_{i,j}\frac{\partial p_i(t_{i,j})}{\partial t}]
\end{equation}

where $P(\overrightarrow{r_j})$ is the pixel magnitude at the $jth$
location (at $\overrightarrow{r_j}$) in the reconstruction area.
$p_i(t_{i,j})$ is the ultrasound pressure signal received at the
$ith$ sensor location. $t_{i,j}$ is the TOF from reconstruction
location $\overrightarrow{r_j}$ to the $ith$ transducer location
calculated from the FMM. In the conventional BP algorithm, $t_{i,j}$
is calculated analytically by assuming a constant speed. It is noted
that, Eq.~\ref{BP} can be viewed as an approximation to the exact BP
algorithm~\cite{HXU2007}. It assumes a circular scanning geometry
and that the scanning radius is considerably larger than the region
of interest~\cite{deanben2011}.

\section{Numerical simulation}
\begin{figure*}[h!]
\includegraphics[scale=0.5]{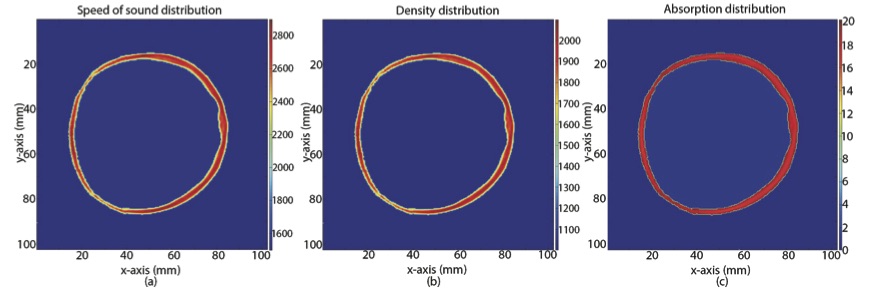}
\caption{(Color online) Acoustic properties derived from the CT scan
for a monkey skull.(a) speed of sound distribution (m/s). (b)
density distribution (kg/m$^3$). (c) absorption distribution
(dB/cm/MHz).}\label{acousticProperties}
\end{figure*}

\begin{figure*}[h!]
\includegraphics[scale=0.17]{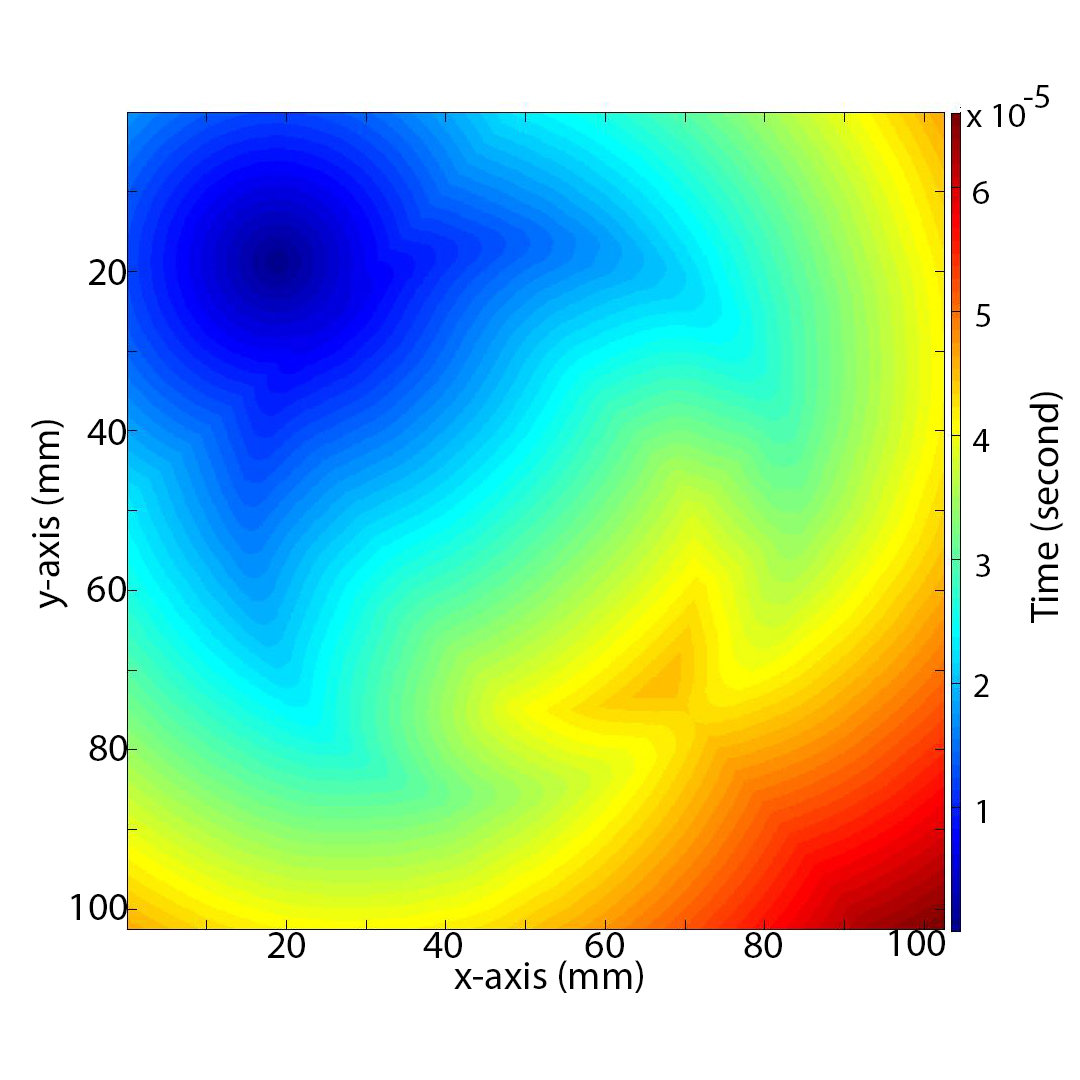}
\caption{(Color online) Arrival times calculated by FMM when a point source is
assigned at a sensor located at (18.36 mm, 18.36 mm). }\label{eikonal_eg}
\end{figure*}

\begin{figure*}[h!]
\includegraphics[scale=0.3]{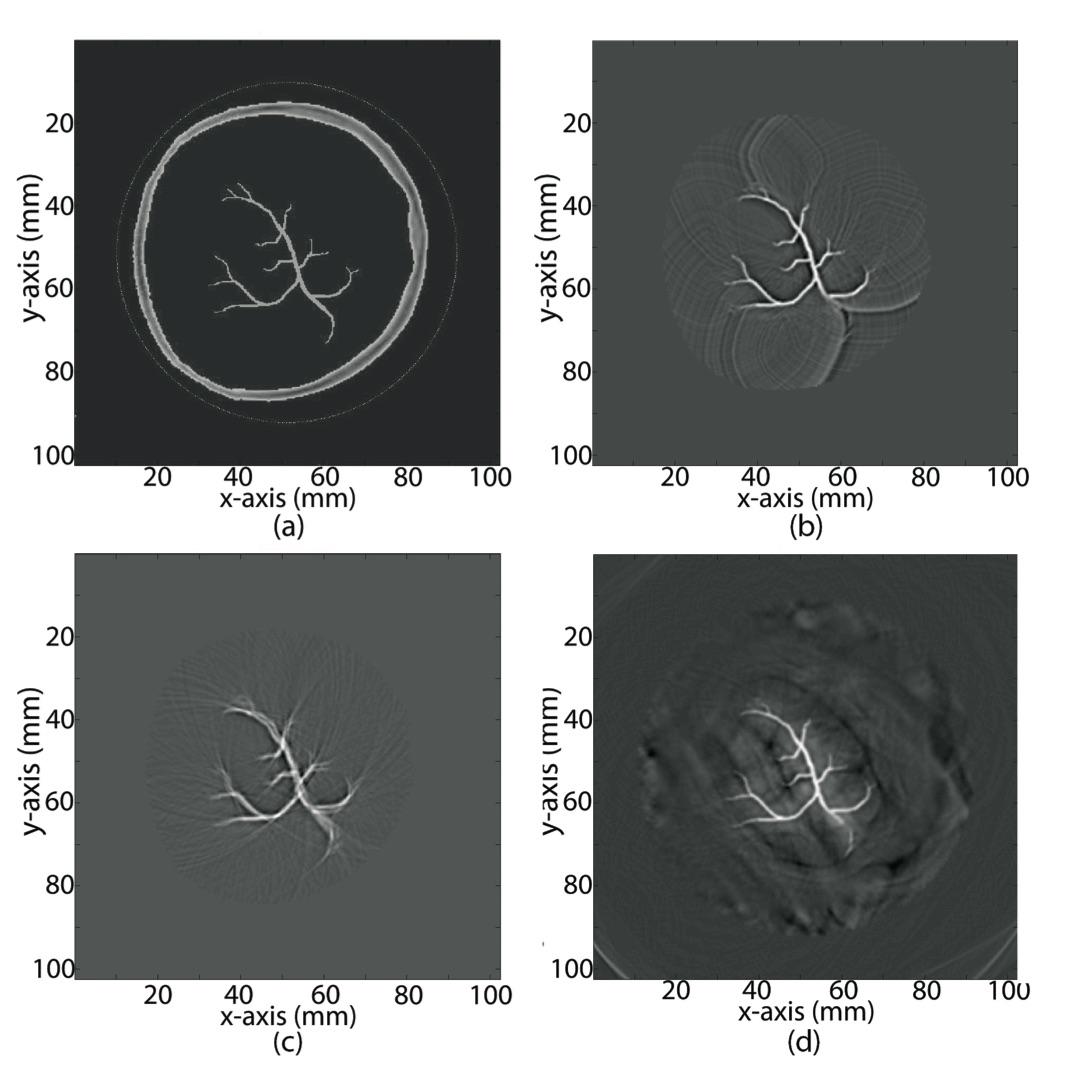}
\caption{(a) Monkey brain model. The white dots encircling the skull
show the locations of the 512 sensors. The cerebral vascular is the
object to be imaged. Reconstructed images using different algorithms
are compared. (b) FMM based BP reconstruction; (c) conventional BP
using the optimal background speed of sound; (d) time reversal
reconstruction. }\label{reconstruction_results}
\end{figure*}

In this section, 2D numerical simulations were implemented to
validate the proposed FMM based BP algorithm. A rhesus macaque
monkey skull was chosen for the numerical simulation. The acoustic
properties (speed of sound, density and attenuation) of the monkey
skull were derived from its CT scan based on the method described in
\cite{AubryJF2003} (Fig.~\ref{acousticProperties}). The sound speed
of the brain tissue was $1540$ m/sec and the mean sound speed of
skull was around $2450$ m/sec.  The absorption was assumed to be
linearly frequency dependent and the attenuation of the brain tissue
was set to be $0.65$ dB/cm/MHz. A ring array of $512$ sensors was
positioned around the skull to receive the acoustic signals
(Fig.~\ref{reconstruction_results} (a)). These sensors were assumed
to have flat frequency response for simplification. For the forward
photoacoustic simulation carried out by K-WAVE, the spatial
resolution was set to be $0.195$ mm. A majority of the broadband
photoacoustic signal energy resides at frequencies less than $1$ MHz
\cite{HuangC2012}, and the resolution was chosen because it is about
$1/8$ of the wavelength at $1$ MHz and was considered sufficiently
small. The Courant-Friedriches-Lewy (CFL) number was $0.03$. This
small CFL number was chosen to maintain the stability of the k-space
algorithm under extremely large absorption in the skull. The end
time for both forward and backward wave propagation was $ 1.5\times
d/c_{0}$, where $d$ is the diameter of the ring array and $c_0$ is
the background speed of sound. Results did not change significantly
when using an end time greater than this. Figure~\ref{eikonal_eg}
presents the TOF map acquired from FMM due to a point source at an
arbitrary sensor point. The arrival time distribution within the
skull region will be used for correcting the phase delay in the BP
algorithm. In this study, FMM was applied to the speed of sound
distribution derived from the CT scan. A recent study also used CT
scans of skulls for phase correction, but the reconstruction was
based on the time reversal reconstruction \cite{HuangC2012}.

An image with the same resolution as in the forward simulation was
reconstructed using the FMM based BP method, as shown in
Fig.\ref{reconstruction_results} (b). For computational efficiency
purpose, the reconstructions were only implemented for the area
within the skull, i.e., $85291$ pixels in total were reconstructed.
The conventional BP reconstruction without phase correction and the
time reversal reconstruction were also implemented for comparison.
For the conventional BP, a constant speed of sound at 1560 m/s was
used and was determined by an automatic sound speed selection
algorithm~\cite{treeby2011}. The results are shown in
Fig.~\ref{reconstruction_results} (c) and (d), respectively. The
image quality resulted from the FMM based BP method
(Fig.~\ref{reconstruction_results} (b)) is considerably better than
that of the conventional BP (Fig.~\ref{reconstruction_results} (c)),
which was distorted and blurred. As the phase correction step (FMM)
in the proposed method only needs to be calculated during the
off-line process once (the entire FMM computation took $184.28$
seconds in this study), the online computation time of the proposed
method is the same as the conventional BP. It is noted that the
off-line process needs to be repeated though if the sensor positions
are changed. To achieve efficient computation, the reconstruction
algorithm was implemented in GPU\cite{wangkun2013} using an NVIDIA
Tesla C2075 with $6$GB of memory and $14$ multiprocessors with
computation capability $2.0$. The GPU variable grid $dim3$ was set
to be $(32,2,1)$ and the GPU variable thread $dim3$ was set to be
$(32,32,1)$. The computation time, including transferring data from
CPU memory to GPU memory, image reconstruction, and transferring
data back to CPU, was $241.83$ ms in total, indicating almost $4$
frames per second (fps). This is about three orders of magnitude
faster than on an Intel Xeon E5620 CPU. A similar amount of
computational speed-up using GPUs was also recently
reported~\cite{wangkun2013}. The processing time can be further
reduced to $123.79$ ms if the spatial resolution was changed from
$0.195$ mm to $0.39$ mm ($21323$ pixels) without compromising the
accuracy. For comparison, the time reversal reconstruction was also
implemented in GPU with the CFL number set to $0.24$ and a spatial
resolution of $0.195$ mm. A larger CFL number was used in the
backward propagation because the attenuation was not considered. The
computation time was $43.89$ seconds.  It is worth pointing out that
with the FMM based BP, a specific region of interest can be selected
for reconstruction to further reduce the reconstruction time. This
is in general not possible for the time reversal reconstruction.

\begin{figure*}[h!]
\includegraphics[scale=0.3]{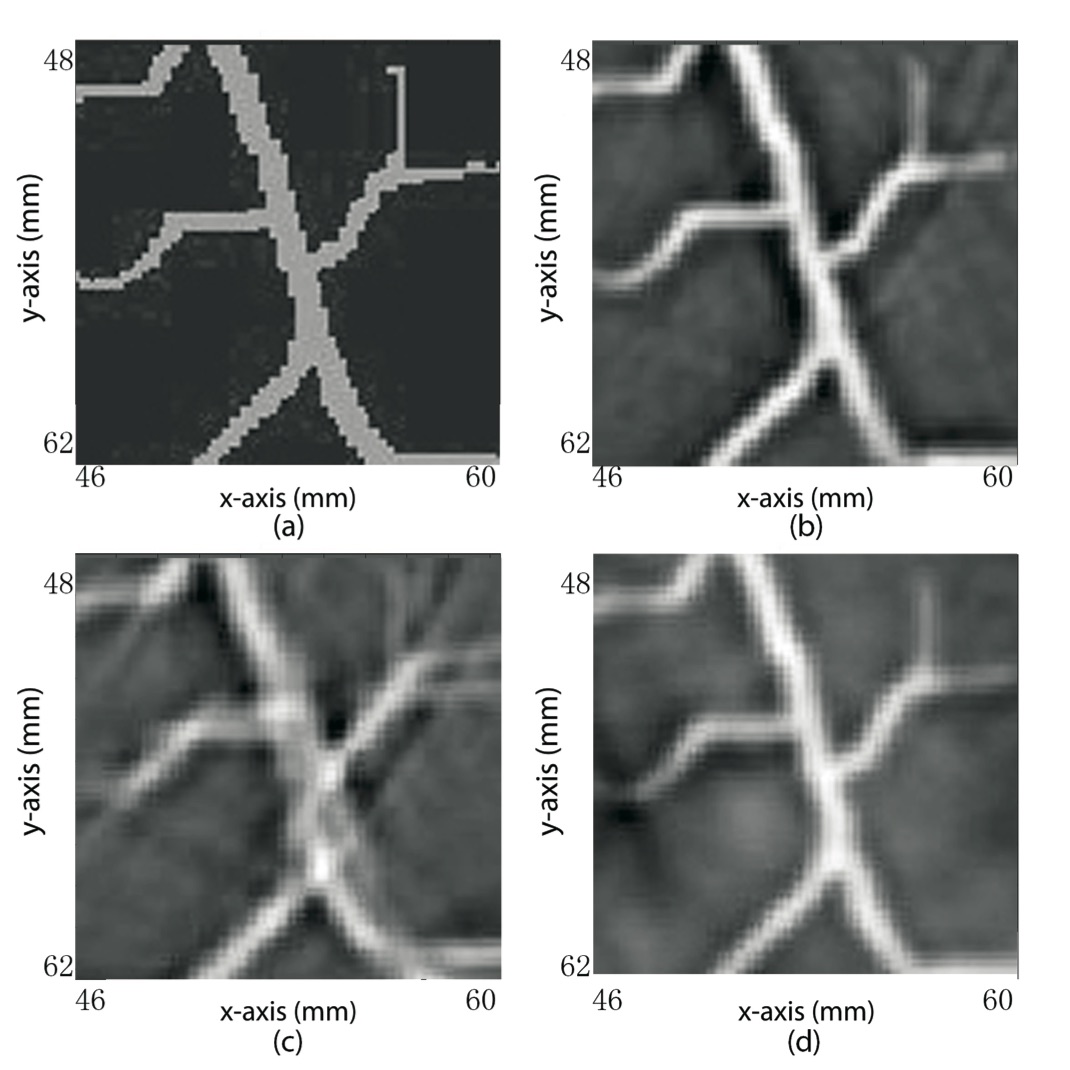}
\caption{Close-up comparison between different algorithms. The
center of the vessel is shown. (a) True image; (b) FMM based BP
reconstruction; (c) conventional BP using the optimal background
speed of sound; (d) time reversal reconstruction.
}\label{comparison}
\end{figure*}

The result of the time reversal method, as shown in
Fig.~\ref{reconstruction_results} (d), has the best accuracy as
expected. Comparing Fig.~\ref{reconstruction_results} (b) to (d), it
is found that the main difference is the region close to the skull,
which is not well reconstructed by the modified BP. This is
consistent with the conclusion of \cite{JinX2008}: the region close
to the skull is more susceptible to the phase aberration from the
skull. Even though the result from the time reversal reconstruction
is less distorted than the result of the proposed method, the time
reversal method is significantly more time consuming as demonstrated
above, therefore is not feasible for real-time imaging. Close-up
comparisons are also provided in Fig.~\ref{comparison}, where the
center of the vessel is shown. It should be borne in mind that,
photo-acoustic waves in 2D are not compactly supported. This means
that time reversal reconstruction for PAT is not exact in 2D.
However, 3D simulations are extremely time- and memory-consuming and
therefore were not investigated.

\begin{figure*}[h!]
\includegraphics[scale=0.4]{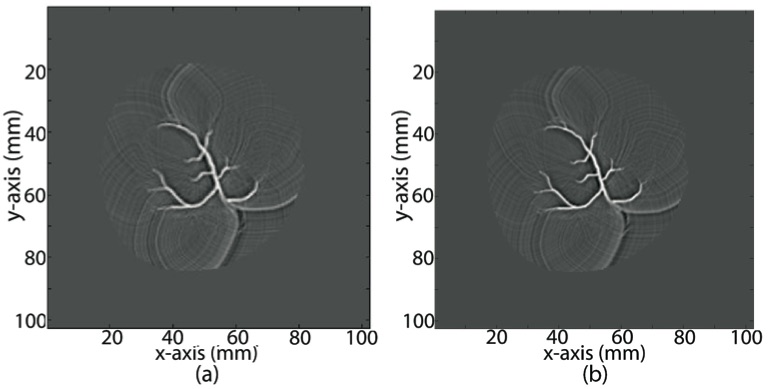}
\caption{(a) Reconstructed image with the speed of sound error. (b)
Reconstructed image without the speed of sound error.
}\label{error_study}
\end{figure*}

Finally, we investigated the robustness of our algorithm against
errors in the skull sound speed. Similar to a previous
study~\cite{HuangC2013}, $1.7\%$ (with respect to maximum value)
uncorrelated Gaussian noise with mean value of $1.7\%$ of the
maximum value was added to the speed of sound maps. The density
errors were not studied as it should not affect the BP algorithm.
Results are shown in Fig.~\ref{error_study} and no significant
changes are observed. The width of the vessel, however, does seem to
be overestimated when speed of sound errors are present. Comparing
Fig.~\ref{error_study} (a) to Fig.\ref{reconstruction_results} (c),
the modified BP algorithm is still more accurate than the original
BP using a constant speed of sound. For example, the top portion of
the vessel seems to split into two parts in
Fig.~\ref{reconstruction_results} (c). This artifact is not seen in
Fig.~\ref{error_study} (a). In this study, the monkey skull was
relatively rounded. We expect the conventional BP to be even more
inaccurate when the skull has a less regular shape (e.g., more or
less elliptical, such as for human skulls) and ring array is used
again. This is because the phase delays in this case become more
non-uniform with regard to each sensor, therefore can not be easily
compensated using a constant phase correction.

\section{Conclusion}
This paper demonstrates a method for correcting the phase aberration
in PAT imaging in heterogeneous media. Even though the skull was
assumed to be the heterogeneity in this study, the proposed
algorithm can be applied to general tissue heterogeneities, such as
in breasts, kidneys, etc. The conventional BP method was modified by
taking the delayed time/phase into account, which was conveniently
computed by the FMM. Improvement over conventional BP algorithm in
terms of reconstructed images was achieved. Compared to the most
accurate time reversal reconstruction, the proposed algorithm was
two orders of magnitude faster, and at the curate stage can be
implemented in quasi-real-time. However, we envision that truly
real-time 2D PAT reconstruction is possible in the near future with
the fast development of GPU technology. We also note that in
practice, BP in fact could have a higher computational complexity
than time reversal reconstruction~\cite{Burgholzer2007}. In this
study, BP was shown to be faster because it can be significantly
accelerated utilizing the superior parallel computing power of
GPUs~\cite{wangkun2013}. It is not the case, though, for time
reversal reconstruction, since the solution at each time step is
dependent on the previous one. In other words, all the time steps
have to be implemented sequentially as opposed to in parallel.

Because the modified BP algorithm has the same on-line computation
speed to the original BP algorithm, we expect such an algorithm,
once implemented for 3D image reconstruction, could be completed on
the order of seconds using currently existing
GPUs~\cite{wangkun2013}.

 As the current study is a proof-of-concept numerical study, we intend to
 implement the algorithm proposed here on ex vivo monkey skulls to further
 evaluate the potential of the approach as a future study. In this way,
 dispersion, shear waves, and out-of plane refraction could be naturally taken
 into account.

\bibliographystyle{./IEEEtran}

\bibliography{document}

\end{document}